\documentclass[aps,prl,12pt,showpacs,showkeys,nobibnotes]{revtex4}

\usepackage{amsmath}    
\usepackage{graphicx}   
\usepackage{verbatim}   
\usepackage{color}      
\usepackage{subfigure}  
\usepackage{hyperref}

\begin{document}
\preprint{}

\title{A simple mathematical formulation of the correspondence principle}
\author{J. Bernal}
\email{jorge.bernal@dacb.ujat.mx}
\affiliation{Universidad Ju\'{a}rez Aut\'{o}noma de Tabasco. Divisi\'{o}n Acad\'{e}mica de 
Ciencias B\'{a}sicas.C.P. 86690. Cunduac\'{a}n, Tabasco,M\'{e}xico.}

\author{Alberto Mart\'{i}n-Ruiz}
\email{alberto.martin@nucleares.unam.mx}
\affiliation{Instituto de Ciencias Nucleares, Universidad Nacional Aut\'{o}noma de 
M\'{e}xico. C.P. 04510 M\'{e}xico, D.F., M\'{e}xico}

\author{J. C. Garc\'{i}a-Melgarejo}
\email{j\_melgarejo@inaoep.mx}
\affiliation{Instituto Nacional de Astrof\'{i}sica, \'{O}ptica y Electr\'{o}nica. INAOE.
C.P.72000, Santa Mar\'{i}a Tonantzintla, Puebla, M\'{e}xico}

\date{\today}

\begin{abstract}

In this paper we suggest a simple mathematical procedure to derive the classical probability density of quantum 
systems via Bohr's correspondence principle.  Using Fourier expansions for the classical and quantum distributions, 
we assume that the Fourier coefficients coincide for the case of large quantum numbers $n$. We illustrate the 
procedure by analyzing the classical limit for the quantum harmonic oscillator, although the method is quite general. 
We find, in an analytical fashion, the classical distribution arising from the quantum one as the zeroth order term 
in an expansion in powers of Planck's constant. We interpret the correction terms as residual quantum effects at the 
microscopic-macroscopic boundary.

\end{abstract}


\maketitle
    
	In physics, a new theory should not only describe phenomena unexplained by the old theory but must also be 
consistent with it in the appropriate limit \cite{A1}. In this sense, newtonian mechanics can be recovered from 
relativistic mechanics in the domain of low velocities compared with the speed of light in the vacuum. Since its 
formulation, quantum mechanics has established itself as the most successful physical theory for the description of 
microscopic systems, such as atoms and elementary particles. Unlike special and general relativity, relations between 
classical and quantum mechanics are more subtle, given that the conceptual framework of these theories are 
fundamentally different. While in classical mechanics it is possible to know the exact position and momentum of a 
particle at any given time, quantum mechanics only specifies the probability of finding a particle at a certain 
position \cite{A2}.	
	
	The first statement of a mathematical procedure to obtain the classical limit of quantum mechanics can be traced 
back to Max Planck \cite{A3}. He postulated that classical results can be recovered from quantum ones when Planck’s 
constant is taken to zero. Planck originally formulated this limit to show that his energy density for black body 
radiation approaches the classical Rayleigh-Jeans energy density when $\hbar \rightarrow 0$. A different approach is 
due to Niels Bohr \cite{A4}. He postulated that the classical behavior of periodic quantum systems can be determined when the principal quantum number is large. Bohr enunciated it in this way because in his model of the hydrogen atom the transition frequency between two neighboring energy levels tends to the classical orbital frequency of the electron when $n\gg1$. Some researchers, however,  have argued that the two methods are not equivalent \cite{A5, A6, A7}.
	
	Textbooks and articles on quantum mechanics usually discuss a variety of ways to make the connection between 
classical and quantum physics. Most of them are based on either Planck’s limit or Bohr’s correspondence principle. 
For example, the WKB \cite{A9, A10, A11} and quantum potential \cite{A12} methods and the phase space formulation of 
quantum mechanics are discussed using Planck's limit, while some authors \cite{A2, A13} compare the classical and 
quantum probability densities for both position and momentum, showing that these distributions approach each other in 
a locally averaged sense (coarse-graining) for large quantum numbers $n$. There are other proposals, like Ehrenfest´s 
Theorem \cite{A8}, based on semi-classical approximations to quantum mechanics. Another method is by means of 
coherent states. The standard coherent states of the one-dimensional harmonic oscillator \cite{S, S1, G} are localized wave packets which follow the classical equations of motion. However, for non-quadratic Hamiltonians this only holds approximately over short times.
	
	Wigner´s phase-space formulation of quantum mechanics offers a comprehensive framework in which quantum phenomena 
can be described using classical language. The Wigner distribution function (WDF), however, does not satisfy the 
conventional properties of a probability distribution \cite{A18}; e.g., WDF is in general positive semi-definite. 
Therefore, in order to interpret it as a classical probability distribution, strictly one needs to restrict the 
analysis to situations where it is non-negative (this is the case for coherent and squeezed vacuum states only) 
\cite{A19, A20}. W. B. Case has made a careful discussion of the classical limit and its difficulties via WDF 
\cite{A21}.
   
	According to Bohr´s correspondence principle, classical mechanics is expected to be valid in the regime in 
which dynamical variables are large compared to the relevant quantum units \cite{B14}. In addition, some authors 
\cite{A2, A13, B22, A23} suggest that we must compare the same physical quantities in both approaches , e.g.  
probability distributions and not trajectories or wave functions.
    
	In 1924, Heisenberg made an attempt to give Bohr's correspondence principle an exact mathematical form in order 
to apply to simple quantum systems. He suggested that for a classical quantity $f(t)$ in the case of large quantum 
numbers, the following approximate relation is valid:   
    
\begin{equation} \label{eq1}
\left\langle \psi _{n+m}\mid f\mid \psi _{n}\right\rangle =\left\langle
n+m\mid f\mid n\right\rangle \exp [i/\hbar (E_{n+m}-E_{n})t]\approx
f_{m}(n)\exp [im\omega (n)t]
\end{equation} 

where $f_{m}(n)$ is the mth Fourier component of the classical variable $f$ and $\omega (n)$ is the classical 
frequency \cite{B24, A25}. The application of this procedure, however, was limited to the study of light polarization 
in atoms subject to resonant fluorescence \cite{WE1, WE2}.

	In 1926, E. Schr{\"o}dinger proposed a different application of the correspondence principle applied to the quantum harmonic oscillator. His approximation consists of adding all the wave function oscillation modes, generating a semiclassical wave packet \cite{ES}, from which other interesting ideas have recently evolved \cite{A26, A27}. On the other hand, discrepancies and discussion remains about the adequacy of Bohr's correspondence principle \cite{A28, 
A29, A30, A31}. Some suggest that the harmonic oscillator does not have a true classical limit when described by 
means of stationary states \cite{B17} and others argue that this system violates Bohr's correspondence principle 
\cite{A32}. 

	In this paper, we suggest a conceptually simple mathematical procedure to connect the classical and quantum
probability densities using Bohr's correspondence principle.

	It is well know that for periodic systems, the quantum probability distribution (QPD) $\rho ^{QM}\left( 	
x,n\right)$ is an oscillatory function, while the classical probability distribution (CPD) $\rho^{CL}(x)$  does 
not have this behavior. However, both functions can be written as a Fourier expansion, i.e.
	
\begin{eqnarray}
\rho ^{QM}\left( x,n\right) =\int f^{QM}\left( p,n\right) e^{i\frac{px}{\hbar }}dp \label{eq2}
\\
\rho ^{CL}\left( x\right) =\int f^{CL}\left( p\right) e^{i\frac{px}{\hbar }}dp \label{eq3}
\end{eqnarray}

where $f^{QM}\left( p,n\right)$ and $f^{CL}\left( p\right)$ are the quantum and classical Fourier coefficients, 
respectively. In addition, we know that for simple periodic systems these distributions approach each other in a 
locally averaged sense for large quantum numbers. This implies that the Fourier expansion coefficients should 
approach each other for $n \gg1$:

\begin{equation} 
f^{QM}\left( p,n\right) \sim f^{CL}\left( p\right) \label{eq4}
\end{equation}

In order to make this comparison we first substitute the value of the principal quantum number $n$ by equating the 
quantum and classical expressions \cite{A2, A13, A23}. Note that the Planck constant keeps a finite value, so 
$\hbar$-dependent corrections may arise, as implied by equation(\ref{eq4}).

	Our proposal can be summarized as follows. First we calculate the coefficients of the expansion $f^{QM}\left( 
p,n\right)$ by using the Fourier transform of QPD, and then obtain its asymptotic behavior for large $n$. We then 
equate the classical and quantum expressions for the energy, to define the value of the principal quantum number. 
Finally calculating the inverse Fourier transform we obtain, at least in a first approximation, the CPD. The 
procedure can be also applied to probability distributions in momentum space.

	We now apply this procedure to the harmonic oscillator \cite{B14, B22}. The QPD for a one-dimensional harmonic 
oscillator is given by

\begin{equation} \label{eq5}
\rho ^{QM}\left( x,n\right) = \sqrt{\frac{\alpha }{\pi }}\frac{1}{2^{n}n!}H_{n}^{2}\left( \sqrt{\alpha }x\right) e^{-\alpha x^{2}}
\end{equation}

where $\alpha = \frac{m\omega }{\hbar }$. One of the main differences between the classical and quantum descriptions 
of the harmonic oscillator is that the QPD is distributed completely throughout the x-axis, while the CPD is bounded 
by the classical amplitude. However, when increase the value of the principal quantum number $n$, the QPD exhibits a 
confinement effect, akin to the classical behavior. 

	We now calculate the Fourier coefficients. The corresponding integral can be found in many handbooks of 
mathematical functions \cite{B33, B34}: 

\begin{equation} \label{eq6}
f^{QM}\left( p,n\right)=e^{-\frac{p^{2}}{4m \omega \hbar}}L_{n}\left( \frac{p^{2}}{%
2m \omega \hbar}\right)  
\end{equation}

where $L_{n}$ is a Laguerre polynomial of degree  $n$. We remark that the mathematical structure of the coefficients 
$f^{QM}\left( p,n\right)$  is similar to the Wigner function for the harmonic oscillator \cite{B35}, but formally 
different, due to the dependence of the wave functions on parity \cite{DD}.Technically, the WDF is a member of the Cohen class of phase-space distributions which is related to the fractional Fourier transform \cite{A36}, and not with the usual Fourier transform as is the case for the expansion coefficients.

	The asymptotic behavior of Fourier coefficients for n large is also well known. Szeg\"{o} \cite{B37} finds the 
following iterative relation:

\begin{eqnarray} \label{eq7}
F\left( u^{2}\right)=e^{-\frac{u^{2}}{2}}L_{n}\left( u^{2}\right) &\sim& J_{0}\left(2\sqrt{N}u\right)
-\frac{\pi }{2} \nonumber \\ 
&& \times\int_{0}^{u}t^{3}F\left(t^{2}\right) \left[ J_{0}\left( 2\sqrt{N}u\right) Y_{0}\left( 2\sqrt{N} t\right) -
J_{0}\left( 2\sqrt{N}t\right) Y_{0}\left(2\sqrt{N}u\right) \right]dt
\end{eqnarray}%

where $J_{0}$ and $Y_{0}$ are the usual Bessel functions of the first and second kind respectively, and $N=n+\frac{1}
{2}$. Szeg\"{o} shows that in $N \rightarrow \infty$ limit the iteration terms are strongly suppressed compared to 
$J_{0}$ Bessel function.

	Using the above relation and ℏ$N=\frac{m\omega x_{0}^{2}}{\hbar}$, we can write the asymptotic expression for the 
Fourier coefficients as follows

\begin{eqnarray} \label{eq8}
f^{QM}\left( p,n\right) &\sim& 
J_{0}\left( \frac{px_{0}}{\hbar }\right) 
-\frac{\pi }{2} \nonumber \\ 
&& \times\int_{0}^{\frac{p}{\sqrt{2m\omega \hbar }}}t^{3}F\left(
t^{2}\right) \left[ J_{0}\left( \frac{px_{0}}{\hbar }\right) Y_{0}\left( 2\sqrt{N}t\right) -J_{0}\left( 
2\sqrt{N}t\right) Y_{0}\left( \frac{px_{0}}{\hbar }\right) \right] dt
\end{eqnarray}%

Finally, we compute the inverse Fourier transform. The first term can be obtained directly, while the iterated terms 
can be written as dimensionless integrals

\begin{equation}\label{eq9}
\rho^{QM}\left( x,n\right) \sim \frac{1}{\pi \sqrt{x_{0}^{2}-x^{2}}}+ \frac{1}{2\pi x_{0}} 
\sum\limits_{k=1}^{\infty}\left(-\frac{\pi }{32}\right)^{k}\left( \frac{\hbar }
{S}\right)^{2k}i_{k}\left(x,x_{0}\right)
\end{equation}

where $S=\pi m \omega x_{0}^{2}$ is the classical action and the $i_{k}=(x,x_{0})$ is the kth dimensionless integral. In particular:

\begin{equation}\label{eq10}
i_{1}\left( x,x_{0}\right) =\int_{-\infty }^{+\infty }d\alpha e^{i\alpha 
\frac{x}{x_{0}}}\int_{0}^{\alpha }\beta ^{3}J_{0}\left( \beta \right) \left[
J_{0}\left( \alpha \right) Y_{0}\left( \beta \right) -J_{0}\left( \beta
\right) Y_{0}\left( \alpha \right) \right] d\beta
\end{equation}

We can also evaluate higher order iterations in a simple fashion \cite{B37}.

	Note that the first term in Eq.~(\ref{eq9}) is $\hbar$-independient and corresponds exactly with the CPD 
\cite{A2, A13}. The remaining terms are proportional to increasing powers of $\frac{\hbar}{S}$, which are very small 
for classical systems, so these terms are strongly suppressed compared with the CPD. A residual oscillatory behavior, 
as observed in the QPD is preserved through the harmonic behavior of the iterated integrals. If we now consider 
Planck´s limit, the classical result is exactly recovered. This, however, is not necessary, as the correction terms 
are very small and seem to reflect a residual quantum behavior at the classical level.		

	A complete agreement of both the position and momentum distribution functions at the classical limit is necessary 
for the theory to recover the classical results in the appropriate energy limit \cite{A38}. In this case, due to 
the symmetry of the harmonic oscillator, the QPD in momentum space can be obtained easily,  so the asymptotic behavior of the QPD for large quantum numbers is given by:	
	
\begin{equation}\label{eq11}
\rho^{QM}\left( p,n\right) \sim \frac{1}{\pi \sqrt{p_{0}^{2}-p^{2}}}+ \frac{1}{2\pi p_{0}} 
\sum\limits_{k=1}^{\infty}\left(-\frac{\pi }{32}\right)^{k}\left( \frac{\hbar }
{S}\right)^{2k}i_{k}\left(p,p_{0}\right)
\end{equation}

where $p_{0}$ is its maximum momentum, $S=\pi \frac{p_{0}^{2}}{m\omega}$ is the classical action and $i_{k}(p,p_{0})$ is the same dimensionless integral defined by Eq.~(\ref{eq10}). 

	Expectation values of physical quantities can be calculated using our previous results and the classical values 
are then recovered, i. e.

\begin{subequations} \begin{eqnarray}
\left\langle \hat{x}^{2}\right\rangle \sim \bar{x}_{CL}^{2} \label{eq12}
\\
\left\langle \hat{p}^{2}\right\rangle \sim \bar{p}_{CL}^{2} \label{eq13}
\\
\left\langle \hat{H}\right\rangle \sim E_{CL} \label{eq14}
\end{eqnarray} 
\end{subequations}

where we have not included the correction terms. These results do not ensure that the time dependence of position and 
momentum operators defined by the Heisenberg equation reduces to the classical equations of motion, due to the fact 
that the classical limit is not a single trajectory, but an ensemble of trajectories.

	To summarize, the classical limit problem has been debated since the birth of quantum theory and is still a 
subject of research.  In this paper, we present a simple mathematical formulation of Bohr's correspondence 
principle. We consider the simplest quantum system, the harmonic oscillator, and obtain exact classical results. 
We believe that this approach illustrates in a clear fashion the difference between Planck´s limit and Bohr´s 
correspondence principle.

	Finally, using this simple procedure we find corrections to the exact classical result as a series in the ratio 
$\frac{\hbar}{S}$, which is very small for classical energies but not zero. It would be interesting to test 
whether this energy dependence could be observed for the case of real quantum systems approaching the 
microscopic-macroscopic boundary.
 
\vspace{1cm}
We thank Alejandro Frank Hoeflich and Jos\'{e} Adri\'{a}n Carbajal Dom\'{\i}nguez for his valuable contribution to 
the fulfillment of this work.

\bibliography{references}

\end{document}